\documentstyle[12pt,times,cite]{article}

\newcommand{\mysection}[1]{\setcounter{equation}{0}\section{#1}}

\newcommand{\mrm}[1]{\mbox{\rm #1}}
\newcommand{\half}{{1\over 2}}
\newcommand{\bla}{\hspace{1cm}}
\newcommand{\beq}{\begin{equation}}
\newcommand{\eeq}{\end{equation}}
\newcommand{\bea}{\begin{eqnarray}}
\newcommand{\eea}{\end{eqnarray}}
\newcommand{\eq}[1]{eq.~(\ref{#1})}
\newcommand{\rfn}[1]{(\ref{#1})}

\newcommand{\sla}[1]{\hspace{-0.1ex}\not\hspace{-0.5ex} #1\hspace{0.1ex}}

\newcommand{\gh}{\frac{g}{4c_W}}
\newcommand{\zbb}{Z\rightarrow b\bar{b}}
\newcommand{\zbbg}{Z\rightarrow b\bar{b}g}
\newcommand{\as}{\alpha_s}
\newcommand{\ap}{\frac{\alpha_s}{\pi}}

\newcommand{\yc}{y_{c}}
\newcommand{\rb}{r_b}
\newcommand{\za}{z_\alpha}
\newcommand{\zb}{z_\beta}
\newcommand{\zg}{z_\gamma}
\newcommand{\be}{\beta}
\newcommand{\el}[1]{^{(#1)}}

\renewcommand{\Re}[1]{\mathop{\mrm{Re}}\left\{ #1 \right\}}

\renewcommand{\titlepage}{\clearpage%
\setcounter{footnote}{0}%
\thispagestyle{empty}\pagestyle{plain}\pagenumbering{arabic}%
\kern1mm
\vskip15mm\normalsize}
\newcommand{\docnum}[1]{\hbox to \hsize{\hskip123mm\hbox{#1}\hss}}
\renewcommand{\date}[1]{\hbox to \hsize{\hskip123mm\hbox{#1}\hss}}

\renewcommand{\title}[1]{\vskip1em\begin{center}\Large\bf#1\end{center}\vskip2.5em}
\renewcommand{\author}[1]{\vskip0.5em{\bf #1}\vskip0.5em}
\newcommand{\inst}[1]{\vskip0.3em{ #1}\vskip0.5em}
\renewcommand{\abstract}{\begin{center}{\bf Abstract}\end{center}\quotation}

\newcommand{\anotfoot}[2]{\vfill\noindent\underline{\hspace{6cm}}
\par\noindent #1) #2}
\newcommand{\anotfootnb}[2]{\par\noindent #1) #2}

\begin{document}
\begin{titlepage}
\docnum{CERN-TH.7419/94}
\docnum{hep-ph/9410258}
\vspace{1cm}
\title{Three-jet production at LEP and the  bottom quark mass}
\begin{center}
\author{Mikhail Bilenky$^{*)}$}
\inst{DESY-IfH, Platanenallee 6, 15738 Zeuthen, Germany}
 \author{Germ\'an Rodrigo}
 \inst{Dept. de F\'{\i}sica Te\`orica, Univ. de Val\`encia,\\
 E-46100 Burjassot (Val\`encia), Spain}
 \inst{and}
\author{Arcadi Santamaria$^{**)}$}
\inst{TH Division, CERN, 1211 Gen\`eve 23, Switzerland}
\end{center}
\vspace{0.5cm}

\begin{abstract}
We consider the possibility of extracting the bottom quark
mass from LEP data.
The inclusive decay rate for $\zbb +\cdots$ is obtained at order
$\as$ by summing up the
one-loop two-parton decay rate to the tree-level three-parton rate.
We calculate the decay width of the $Z$-boson
into two and three jets containing the
$b$-quark  including complete quark mass effects.
In particular, we give analytic results for
a slight modification of the JADE clustering algorithm.
We also study the
angular distribution with respect
to the angle formed between the gluon and the quark jets,
which has a strong dependence on the quark mass.
The impact of higher order QCD corrections on these observables
is briefly discussed.
Finally, we present numerical results for some popular
jet-clustering algorithms and show that, indeed,
these three-jet observables are
very sensitive to the $b$-quark mass and well suited for its
determination at LEP.
\end{abstract}

\vspace{0.5cm}
\vfill\noindent
CERN-TH.7419/94\\
\\ \today
\anotfoot{ *}{On leave of absence from
the Joint Institute for Nuclear Research, Dubna, Russia.}
\anotfootnb{**}{On leave of absence from Departament de
F\'{\i}sica Te\`orica,
Universitat de Val\`encia, and IFIC, Val\`encia, Spain.}
\end{titlepage}
\setcounter{page}{1}

\mysection{Introduction}

In the Standard Model of electroweak interactions
all fermion masses are free parameters
and their origin, although linked to the spontaneous symmetry
breaking mechanism, remains secret.
Masses of charged leptons are well measured experimentally and
neutrino masses, if they exist, are also bounded.
In the case of quarks the situation is more complicated because free
quarks are not observed in nature.
Therefore,
one can only get some indirect information on the values
of the quark masses. For light quarks
($m_q < 1$~GeV, the scale at which QCD interactions become
strong), that is, for $u$-,$d$- and $s$-quarks, one can
define the quark masses as the parameters of the Lagrangian that break
explicitly the chiral symmetry of the massless QCD Lagrangian. Then,
these masses can be extracted from a careful analysis of meson
spectra and meson decay constants. For heavy quarks ($c$- and $b$-quarks)
one can obtain the quark masses from the known spectra of the hadronic
bound states by using, e.g., QCD sum rules or lattice calculations. However,
since the strong gauge coupling constant
is still large at the scale of heavy
quark masses, these calculations are plagued by uncertainties and
nonperturbative effects.

It would be very interesting to have some experimental
information on the quark masses obtained at much larger scales
where a perturbative quark mass definition can be used and, presumably,
non-perturbative effects are negligible.
The measurements at LEP will combine
this requirement with very high experimental statistics.

The effects of quark masses can be neglected for many observables in LEP
studies,
as usually quark masses appear in the ratio
$m_q^2/m_Z^2$. For the bottom quark, the heaviest quark produced at
LEP, and taking
a $b$-quark mass of about 5~GeV this ratio is $0.003$, even
if the coefficient in front is 10 we get a correction of about 3\%.
Effects of this order are measurable at LEP, however, as we will see
later, in many cases the actual mass that should be used in the
calculations is the {\it running} mass of the $b$-quark computed at
the $m_Z$ scale: $\bar{m}_b(m_Z) \approx 3$~GeV rendering the effect
below the LEP precision for most of the observables.

While this argument
is correct for total cross sections for production of $b$-quarks it
is not completely true for quantities that depend on other variables.
In particular it is not true for jet cross sections which depend
on a new variable, $\yc$ (the jet-resolution parameter
that defines the jet multiplicity) and
which introduces a new scale in the analysis, $E_c=m_Z\sqrt{\yc}$. Then,
for small values of $\yc$ there could be contributions coming like
$m_b^2/E_c^2 = (m_b/m_Z)^2 /\yc$ which could enhance the mass effect
considerably. In addition mass effects could also be enhanced by
logarithms of the mass. For instance, the ratio of the phase space
for two massive quarks and a gluon to the phase
space for three massless
particles is $1+8 (m_q/m_Z)^2 \log(m_q/m_Z)$.
This represents a 7\% effect
for $m_q=5$~GeV and a 3\% effect for $m_q=3$~GeV.

The high precision achieved at LEP makes these effects relevant.
In fact, they have to be taken into account in the
test of the flavour independence of $\as(m_Z)$
\cite{l3,delphi,opal,aleph,chrin}.
In particular it has been shown \cite{juano} that the biggest systematic
error in the measurement of $\alpha_s^b(m_Z)$ ($\alpha_s$ obtained
from $b\bar{b}$-production at LEP from the ratio of three to two jets)
comes from the uncertainties in the estimate of the quark mass effects.
This in turn means that mass effects have already been seen.
Now one can reverse the question
and ask about the possibility of measuring the mass of the bottom
quark, $m_b$, at LEP by assuming the flavour universality of the strong
interactions.

Such a measurement
will also allow to check the running of
$\bar{m}_b(\mu)$ from $\mu=m_b$ to $\mu=m_Z$ as has been done before
for $\alpha_s(\mu)$. In addition
$\bar{m}_b(m_Z)$ is the crucial input parameter
in the analysis of the unification of
Yukawa couplings predicted by many grand unified theories
and which has attracted much attention in the last years \cite{guts}.

The importance of quark mass effects in $Z$-boson decays
has already been discussed in the literature \cite{rev}.
The complete order $\as$ results for
the inclusive decay rate of $Z\rightarrow b\bar{b}+b\bar{b} g+\cdots$
can be found\footnote{The order $\as$ corrections
to the vector part, including
the complete mass dependences, were already known from QED calculations
\cite{schwinger}.} in \cite{complas}.
The leading quark mass effects
for the inclusive $Z$-width are known
to order $\as^3$ for the vector part
\cite{kuhn1}
 and to order $\as^2$ for the axial-vector part
\cite{kuhn2}.
Quark mass effects for three-jet final states
in the process
$e^+e^- \rightarrow q\bar{q}g$
were
considered first in \cite{ioffe} for the photonic channel and extended
later to the $Z$ channel in \cite{rizzonilles} and \cite{zerwas}.
Recently \cite{ballestrero} calculations of the three-jet
event rates,
including mass effects, were done for the most popular jet
clustering algorithms using the Monte Carlo approach.

In this paper we will discuss the possibility of measuring
the $b$-quark mass at LEP, in particular,
we study bottom quark mass effects in $Z$ decays into
two and three jets. In section 2 we calculate the inclusive decay
rate $Z\rightarrow b\bar{b}+b\bar{b} g+\cdots$ at order $\alpha_s$ by
summing one-loop virtual corrections to $Z\rightarrow b\bar{b}$ and
the real gluon bremsstrahlung contribution.
Dimensional continuation is used to regularize both infrared (IR) and
ultraviolet (UV) divergences.
Phase space integrations are also done in $D$ dimensions.
This calculation  allows us
to understand the details of the cancellation of IR divergences and
how some, potentially large,
logarithms of the quark mass are absorbed in
the running quark mass $\bar{m}_b(m_Z)$. In section 3 we calculate
analytically the two and three-jet event rates in terms of the
jet-resolution parameter $\yc$ and the mass of the quark for a slight
modification of the well-known JADE algorithm \cite{jade} suitable for analytic
calculations with massive quarks. We also present
numerical results for this scheme and for some
of the most popular jet-clustering algorithms (DURHAM ($K_T$),
JADE and E), estimate higher order
contributions and compare with experimental
results obtained by the DELPHI Collaboration
\cite{delphi} for 1990-1991 data. If the gluon
jet can be identified with good efficiency a very interesting observable,
which strongly depends on the quark mass,
is the angular distribution with
respect to the angle formed between the quark  and the gluon jets.
This distribution is calculated for massless quarks in section 4:
analytically for JADE-type algorithms and numerically for the DURHAM
algorithm. We also compute numerically the ratio of
massive to massless angular distributions for the four jet-clustering
 algorithms.
In section 5 we summarize the results obtained in the paper
and comment on the possibility of
using them to measure the $b$-quark mass in LEP experiments. Finally
in the four appendices we collect all the functions and formulae needed in
the body of the paper.

\mysection{The inclusive decay rate $\zbb$}

The main purpose of this paper is to investigate $b$-quark mass effects
in $Z$ decays into two and three jets. Since at order $\as$
the inclusive decay rate $Z \rightarrow b\bar{b}+\cdots$ is given by the
sum of the two- and three-jet decay widths we will start
by studying this quantity.

To calculate the total decay rate to order $\as$ one has
to sum up the virtual one-loop gluonic corrections to the
$Z \rightarrow b \bar{b}$ with the real gluon bremsstrahlung.
Both
contributions are separately  infrared
divergent for massless gluons, therefore,
some regularization method for the IR divergences is needed.
The sum is, however, IR finite.

Since there are many subtleties in this calculation, we sketch it
in this section.
Both processes, $\zbb$ at one loop and $\zbbg$, are calculated
in arbitrary
dimension $D=4-2\epsilon$ and dimensional regularization is
used to regularize the IR divergences \cite{dimreg}.
At order $\as$ and for massive quarks all IR divergences appear as
simple poles $1/\epsilon$.
We show how the the divergences cancel in the sum
and obtain the total inclusive rate.
\par
The first step is to compute the decay width $Z\rightarrow  b \bar{b}$ at
tree-level in dimension $D$. Since
there are no IR divergences in this case
it is not necessary to do the calculations in arbitrary space-time dimensions.
However, there are IR divergences at the
one-loop level and  $\epsilon$ factors could lead to finite
contributions when
multiplied by the divergent terms.
\par
The amplitude for the decay $\zbb$ in $D$ dimensions is
\begin{equation}
T_{b}^{(0)} = \mu^\epsilon \gh
\bar{u}_1 \gamma_\mu (g_V +g_A\gamma_5) v_2
\epsilon^\mu(q)~,
\label{amp0}
\end{equation}
where  the factor $\mu^\epsilon$ has been included to make the
gauge weak coupling $g$
dimensionless in $D$ dimensions;
$u_1$ and $v_2$ are short-hand
notations for the quark (antiquark)
spinors, $u_1=u(p_1)$ and $v_2=v(p_2)$,
$\epsilon^\mu(q)$ stands for the polarization vector of the $Z$-boson and
$g_V$ ($g_A$) are the vector (axial-vector) neutral current
couplings of the quarks in the
Standard Model. At tree level and for the $b$-quark we have
\begin{equation}
g_V=-1+\frac{4}{3} s^2_W~, \bla
g_A= 1~.
\end{equation}
Here we denote by
$c_W$ and $s_W$ the cosine and the sine of
the weak mixing angle.
\par
Taking the square of the amplitude, averaging over initial state polarizations,
summing  over final
state polarizations, and adding the phase space factor for the two-body
decay given in appendix~A~ \cite{dimreg} we obtain the following
decay width in $D$ dimensions,
\begin{equation}
\Gamma_{b}^{(0)} = C_{b} A_{b} \beta^{1-2\epsilon}~,
\label{gammab0}
\end{equation}
with
\begin{equation}
\label{cbb}
C_{b} =m_Z  \frac{g^2 }{c_W^2 64\pi}
\frac{\Gamma(1-\epsilon)}{\Gamma(2-2\epsilon)}
\left(\frac{m_Z^2}{4\pi\mu^2}\right)^{-\epsilon}~,
\end{equation}
and
\begin{equation}
\label{A0}
A_{b} =
\frac{1}{2}(3-\beta^2-2\epsilon) g_V^2+\beta^2(1-\epsilon) g_A^2~.
\end{equation}
In these expressions $\beta$ is the relative velocity of the produced
quarks
\begin{equation}
\beta= \sqrt{1-4r_b}~, \bla r_b=\frac{m_b^2}{m_Z^2}~.
\end{equation}
\par
At the one-loop level (see diagrams in fig.~\ref{feynman}b),
and after renormalization of the UV
divergences\footnote{Note that conserved currents or partially conserved
currents as the vector and axial currents do not get renormalized.
Therefore, all UV divergences cancel when one sums properly self-energy
and vertex diagrams. The remaining poles in $\epsilon$ correspond to
IR divergences. One can see this by separating carefully the poles
corresponding to UV divergences from the poles corresponding to IR
divergences.}, the amplitude can be conveniently parameterized in terms of
three form factors, $f_V$, $f_A$ and $f_T$,
\begin{equation}
\label{amp1}
T_{b} = \mu^\epsilon \gh
\bar{u}_1 \left(g_V\left((1+\half C_g f_V)\gamma_\mu+
i \half C_g f_T \frac{\sigma_{\mu\nu} q^\nu}{2m_b}\right)
 +g_A(1+\half C_g f_A) \gamma_\mu\gamma_5\right) v_2
\epsilon^\mu(q)~,
\end{equation}
where $C_g$ is defined as follows,
\begin{equation}
C_g=\frac{\alpha_s}{\pi} \left(\frac{m_Z^2}{4\pi\mu^2}\right)^{-\epsilon}
\frac{1}{\Gamma(1-\epsilon)}~.
\end{equation}
Here and below we will conventionally use $\as=\as(m_Z)$ to denote the value
of the running strong coupling at the $m_Z$-scale.
\par
The form factors, $f_V$, $f_A$ and $f_T$, are related by
\begin{equation}
\label{frelation}
f_V=f_A+f_T~.
\end{equation}
The two functions, $f_V$ and $f_A$,
contain an IR divergence,
while $f_T$ is
finite. Separating the divergent parts, we can rewrite the real parts
of the form factors as follows
(at order $\as$ the imaginary  parts will not contribute)
\begin{eqnarray}
\Re{f_V} &=& -\frac{1}{\epsilon} f_\epsilon + f_{Vf}~,\\
\Re{f_A} &=& -\frac{1}{\epsilon} f_\epsilon + f_{Af}~,\\
\Re{f_T} &\equiv& f_{Tf} ~,
\end{eqnarray}
where all functions $f_\epsilon$, $f_{Vf}$, $f_{Af}$ and $f_{Tf}$
are given in
appendix~B.
Note that, as expected, the IR divergent part of the amplitude
is proportional to the tree-level amplitude \eq{amp0}.
As the IR divergence manifests itself as a single pole in $\epsilon$, clearly,
we only need to keep everywhere
terms linear in $\epsilon$.
\par
  From  the  amplitude \rfn{amp1}
we obtain the one-loop corrected width in $D$ dimensions
\[
\Gamma^D_{b} = \Gamma_{b}^{(0)} + \Gamma_{b}^{(1)}~,
\]
with
\begin{equation}
\label{gammab1}
\Gamma_{b}^{(1)} =  -C_g f_\epsilon \frac{1}{\epsilon} \Gamma_{b}^{(0)}+
C_b C_g  (g_V^2 F_V+g_A^2 F_A)~,
\end{equation}
where the finite functions $F_V$ and $F_A$ are
given in appendix~B in terms of the form factors and
 $\Gamma_{b}^{(0)}$ is given by \eq{gammab0}
\par
The $O(\as)$ result, \eq{gammab1}, is
divergent for $\epsilon \rightarrow 0$ because
the IR divergences associated with massless gluons running in the loops.
To get a finite
answer at this order we also need to include
gluon bremsstrahlung from the quarks. This has to be computed
by working in $D$ dimensions.
\par
The amplitude for the process $\zbbg$ (the two corresponding diagrams
are given in fig.~\ref{feynman}c)
can be written as
\bea
T_{bg} & = &  \mu^{2\epsilon} \gh g_s~
\bar{u}_1
\left(
\frac{\gamma_\nu (\sla{p}_1+\sla{k}+m_b)\gamma_\mu(g_V+g_A \gamma_5)}
{2(p_1 k)}
\right. \nonumber \\
& + & \left.
\frac{\gamma_\mu (g_V+g_A \gamma_5)(-\sla{p}_2-\sla{k}+m_b)\gamma_\nu}
{2(p_2 k)}
\right)
\frac{\lambda^a}{2} v_2 ~
\epsilon_a^\nu(k) \epsilon^\mu(q)~.
\eea
Here $\lambda^a$ are the
Gell-Mann $SU(3)$ matrices, and
$\epsilon_a^\nu(k)$
is the gluon polarization vector.
\par
The square of the amplitude, in dimension $D$, gives a rather involved
expression that can be conveniently simplified when one realizes that
the most divergent part of it factorizes completely,
even in $D$ dimensions,
due to the factorization theorems for soft and collinear divergences.
\par
Adding the three-body phase space (see appendix~A) we find that the decay
width of $Z\rightarrow b\bar{b} g$ in $D$ dimensions can be written as
\begin{equation}
\label{gammabbg}
\Gamma_{bg} = C_{b} C_g C_F\int d y_1 d y_2 \theta(h_p) h_p^{-\epsilon}
A_{bg}~,
\end{equation}
where $C_F = 4/3$ is the $SU(3)$ group factor, $y_1$ and $y_2$ are
defined in terms of the energy fractions of the two outgoing quarks
\begin{equation}
y_1 = 2 (p_1 k) /m_Z^2 =1-2E_2/m_Z~,
\bla y_2= 2 (p_2 k)/m_Z^2 =1-2E_1/m_Z~
\end{equation}
and $A_{bg}$ comes from the square of the matrix element,
\begin{equation}
\label{ag}
A_{bg} = A_{b} \frac{h_p}{y_1^2 y_2^2}+ g_V^2 h_V+ g_A^2 h_A~.
\label{abg}
\end{equation}
Here $A_{b}$ is the same combination of couplings and masses that
appears in the tree-level decay width to two quarks, \eq{A0},
and the function $h_p$
is given by
\begin{equation}
h_p = y_1 y_2 (1-y_1-y_2)-\rb (y_1+y_2)^2~,
\end{equation}
and
it is exactly the same function that defines the phase space available
for the three-body decay (see \eq{gammabbg} and appendix~A). After
phase space integration this term will contain an IR divergence which
comes from the singularity at $y_1=y_2=0$.
\par
The  functions $h_V$ and $h_A$ describe the vector and the axial-vector
parts of the remainder of the
square of the amplitude which do not generate any IR divergence.
In the limit $\epsilon=0$ they are given by:
\begin{eqnarray}
h_V & = & \frac{1}{2}\left(\frac{y_2}{y_1}+\frac{y_1}{y_2}\right)~,\\
h_A & = & (1+2\rb )h_V +2\rb ~.
\end{eqnarray}
\par
To perform the phase space integration it
is convenient to change variables as follows
\begin{eqnarray*}
y_1 &=& g(z) w~,\\
y_2 &=& g(z) z w~,
\label{chvar}
\end{eqnarray*}
with
\begin{equation}
\label{defg}
g(z)= \frac{z-\rb (1+z)^2}{z(1+z)} = \frac{1}{(1+c)^2}
\frac{(z-c)(1-c z)}{z(1+z)}~
\end{equation}
and
\begin{equation}
\label{defc}
c = \frac{1-\beta}{1+\beta}~.
\end{equation}
Then,  both $h_V$ and $h_A$ only depend on the variable $z$, and the
function $h_p$, which defines phase space and appears explicitly in
\eq{ag}, factorizes completely
\begin{equation}
h_p = g(z)^3 z (1+z) w^2 (1-w)~.
\end{equation}
The function $g(z)$ has zeros at $z_1=c$ and $z_2 = 1/c$.
As phase space is defined by $h_p > 0$ we obtain that the phase space
in terms of the new variables is given by
\begin{equation}
c < z < 1/c\bla \mrm{and}\bla 0 < w < 1~.
\end{equation}
After this change of variables
\eq{gammabbg} can we rewritten as
\begin{equation}
\label{gammabbg2}
\Gamma_{bg} = C_{b} C_g C_F
\int_{c}^{1/c} d z g(z)^2 \int_0^1 d w w h_p^{-\epsilon} A_{bg}~.
\end{equation}
Now the $w$ integration is very simple and leads
to Beta  functions. For the integration of the term of the
amplitude proportional to $A_{b}$ (see \eq{ag}) we get
\begin{equation}
\int_{c}^{1/c} d z  \int_0^1 d w  h_p^{1-\epsilon}\frac{1}{g(z)^2 z^2 w^3}=
B(-2\epsilon,2-\epsilon)  \int_{c}^{1/c} d z
\frac{1}{g(z)^2 z^2}
\left(g(z)^3 z (1+z)\right)^{1-\epsilon}~,
\end{equation}
where the
function $B(-2\epsilon,2-\epsilon)$ has a single pole in
$\epsilon=0$.
In this way, all the divergent behaviour has been factorized  in the
Beta function. Then, to perform the $z$ integration we can expand
the integrant for small $\epsilon$ and keep only terms linear in
$\epsilon$.  The integrations can be easily performed  and the results
written  in terms of logarithms and dilogarithmic functions.
The rest of the integrals do not lead to any divergence and
can be done, without problem, putting $\epsilon$ equal to zero.
\par
After phase space integration, the decay width for $\zbbg$
can be written in the
following form
\begin{equation}
\label{brem}
\Gamma_{bg} =  C_g f_\epsilon \frac{1}{\epsilon} \Gamma_{b}^{(0)}+
C_b C_g \left( g_V^2 G_V +  g_A^2 G_A \right)~,
\end{equation}
where the first term contains the IR divergent part and the IR finite
functions $G_V$ and $G_A$ are given in appendix~B.

The IR divergent part of \eq{brem}
is identical, but with reversed sign, to the  one obtained for
$\Gamma^{(1)}_{b}$, therefore in the sum they will cancel,
as it should be:
\begin{equation}
\label{inclusive}
\Gamma_{b}= \Gamma_{b}^{(0)}+
\Gamma_{b}^{(1)} + \Gamma_{bg} =  \Gamma_{b}^{(0)}+
C_b C_g \left(g_V^2 T_V+g_A^2 T_A\right)~,
\end{equation}
with
\begin{eqnarray}
T_V &=& F_V+G_V~,\\
T_A &=& F_A+G_A~.
\end{eqnarray}
 From the results of the appendix~B we can easily obtain the limit of
these functions for small quark masses,
$m_b \ll m_Z$ ($\rb \ll 1$)
\begin{eqnarray}
T_V &\approx& 1+12\rb,\\
T_A &\approx& 1-6\rb (2\log \rb +1)~.
\end{eqnarray}
If we plug this result into \eq{inclusive} we obtain the well-known result
\cite{kuhn1}
\begin{equation}
\label{inclusiveL}
\Gamma_{b} =
m_Z \frac{g^2}{c_W^2 64 \pi} \left[
g_V^2  \left(
1+\frac{\as}{\pi}(1+12\rb)
\right) +
g_A^2 \left(
1-6 \rb + \frac{\as}{\pi}\left(1-6\rb (2\log\rb+1)\right)
\right)
\right]~.
\end{equation}
It is interesting to note the presence
of the large logarithm, $\log(m_b^2/m_Z^2)$, proportional to
the quark mass in the axial part of the QCD corrected
width, \eq{inclusiveL}.
The mass that appears in all above calculations should
be interpreted as
the perturbative {\it pole} mass of the quark. But in principle the
expression \rfn{inclusiveL} could also be
written in terms of the so-called {\it running} quark mass at the
$m_Z$ scale by using
\begin{equation}
\label{poltorunning}
m_b^2 = \bar{m}_b^2(m_Z) \left[1+2\frac{\as}{\pi}
\left(\log\left(\frac{m_Z^2}{m_b^2}\right)+\frac{4}{3}\right)\right]~.
\end{equation}
Then, we see that all large logarithms are absorbed
in the running of the quark mass from the $m_b$ scale to the $m_Z$ scale
\cite{kuhn1}
and we have
\begin{equation}
\label{inclusiveL2}
\Gamma_{b} =
m_Z \frac{g^2}{c_W^2 64\pi} \left[
g_V^2  \left(
1+\frac{\as}{\pi}(1+12\bar{r}_b)\right)
+
g_A^2  \left(
1-6\bar{r}_b +
\frac{\as}{\pi}(1-22\bar{r}_b)
\right)
\right]~,
\end{equation}
where $\bar{r}_b = \bar{m}_b^2(m_Z)/m_Z^2$.
\par
This result means
that the bulk of the QCD corrections depending on the mass
could be accounted for by using tree-level expressions for the decay
width
but interpreting the quark mass
as the running mass at the $m_Z$ scale.
On the other
hand, since $\bar{m}_b(m_Z) \approx 3$~GeV is much smaller than the
pole mass, $m_b \approx 5$~GeV, it is clear that the quark mass
corrections are much smaller than expected from the na\"{\i}ve use
of the tree-level result with  $m_b \approx 5$~GeV, which would give mass
corrections at the 1.8\% level while in fact, once QCD corrections are
taken into account, the mass corrections are only at the 0.7\% level.
\par
The final results of this section are well known
but we find they could illuminate
the discussion of mass effects in the two- and three-jet event rates
and in the angular distribution with respect to the angle formed
between the quark and gluon jets. Moreover the intermediate results
of this section will be used in the rest of the paper.

\mysection{Two- and three-jet event rates}

According to our current understanding of the strong interactions,
coloured partons, produced  in hard processes,
are hadronized and, at experiment, one only observes colourless particles.
It is known empirically that, in
high energy collision, final particles group in several
clusters by forming  energetic jets,
which are related to the primordial partons.
Thus, in order to compare theoretical predictions
with  experiments, it is necessary
to define precisely what is a jet in both, parton level calculations
and experimental measurements.

As we have seen in the previous section, at order $\as$, the decay
widths of $Z$ into both two and three partons are IR divergent.
The two-parton decay rate is divergent
due to the massless gluons running in the loops.
The $Z$-boson decay width into three-partons has an IR divergence
because
massless gluons could be radiated with zero energy.
The sum, however, is IR finite.
Then it is clear that at the parton-level one can define
an IR finite {\it two-jet decay rate},
by summing the two-parton decay rate
and the IR divergent part of the three-parton decay width, e.g.
integrated over the part of the phase space which contains soft
gluon emission \cite{sw}. The integral over the rest of the phase
space will give the {\it three-jet decay rate}.
Thus we need
to introduce a ``resolution parameter'' in the theoretical
calculations in order to define  IR-safe observables.
Obviously, the resolution parameter, which defines the two- and
the three-jet parts of the three-parton
phase space should be related to the one
used in the process of building jets from real particles.

In the last years the most popular definitions of jets
are based on the so-called jet clustering algorithms.
These algorithms can be applied at the parton
level in the theoretical calculations and also to the
bunch of real particles observed at experiment.
It has been shown that, for some of the algorithms,
the passage from partons to hadrons (hadronization)
does not change much the behaviour of the observables \cite{bethke},
thus allowing to compare theoretical predictions
with experimental results. In what follows we will use the word
particles for both partons and real particles.
\par
In the jet-clustering algorithms jets are defined as follows:
starting from  a bunch of particles
with momenta $p_i$ one computes, for example, a quantity like
\[
y_{ij}=2 \frac{E_i E_j}{s} (1-\cos \theta_{ij})
\]
for all pairs $(i,~j)$ of particles. Then one takes the minimum of
all $y_{ij}$ and if it satisfies that it is smaller than a given quantity
$\yc$ (the resolution parameter, y-cut)  the two particles
which define this $y_{ij}$ are regarded
as belonging to the same jet, therefore, they are recombined into a new
pseudoparticle by defining the four-momentum of the pseudoparticle according
to some rule, for example
\[
p_k = p_i + p_j~.
\]
After this first step one  has a bunch of pseudoparticles and
the algorithm can be applied again and again until all the pseudoparticles
satisfy  $y_{ij} > \yc$.
The number of pseudoparticles found in the end
is the number of jets in the event.
\par
Of course, with such a jet definition the number of jets found in an
event and its whole topology will depend on the value of
$\yc$.
For a given event, larger
values of $\yc$ will result in a smaller number of jets.
In theoretical calculations one can define
cross sections or decay widths into jets as a function of $\yc$,
which are
computed at the parton level, by following exactly the same algorithm.
This procedure leads automatically to IR finite quantities
because one excludes the regions of phase space that cause trouble.
The success of the jet-clustering algorithms is due, mainly,
to the fact that the cross sections
obtained after the hadronization process agree quite well with
the cross-sections calculated at the parton level when the
same clustering algorithm is used in both theoretical predictions
and experimental analyses.
\par
There are different successful jet-clustering algorithms and we refer
to refs.~\cite{bethke,yb} for a detailed discussion and comparison of
these algorithms in the case of  massless quarks.
\par
In the rest of the paper we will use the four jet-clustering algorithms
listed in the table~\ref{table1}, where $\sqrt{s}$ is the total centre of mass
energy.
In addition to the well-known JADE, E and DURHAM algorithms we will
use a slight modification of the JADE scheme particularly useful for
analytical calculations with massive quarks. It is defined by the two
following equations
\[
y_{ij} = 2 \frac{p_i p_j}{s}
\]
and
\[
p_k= p_i+p_j
\]
We will denote this algorithm as the EM scheme.
For massless particles and at the lowest order E, JADE and EM give
the same answers. However already at order $\as^2$ they give different
answers since after the first recombination the
pseudoparticles are not massless anymore and the resolution functions
are different.
\def\arraystretch{1.5}
\begin{table}
\begin{center}
\begin{tabular}{||l|l|l||}
\hline
Algorithm & Resolution & Combination \\
\hline\hline
EM & $2(p_i p_j)/s$ & $p_k= p_i+p_j$ \\
JADE & $2(E_i E_j)/s\ (1-\cos \vartheta_{ij})$ & $p_k= p_i+p_j$ \\
E & $(p_i+p_j)^2/s$ & $p_k = p_i+p_j$ \\
DURHAM  & $2 \min(E_i^2,E_j^2)/s\ (1-\cos \vartheta_{ij})\ \ $ &
$p_k = p_i+p_j$\\
\hline
\end{tabular}
\end{center}
\caption{The jet-clustering algorithms}
\label{table1}
\end{table}
\par
For massive quarks the three algorithms, E, JADE and EM are
already different at order $\as$. The DURHAM ($K_T$) algorithm,
which has been  recently considered in order to avoid exponentiation
problems present in the JADE algorithm \cite{durham,bethke},
is of course completely different from the other
algorithms we use, both in the massive and the massless cases.
\par
In figure~\ref{phase} we plotted the phase-space for two values of $\yc$
($\yc=0.04$ and $\yc=0.14$) for all four schemes (the solid line
defines the whole phase space for $Z \rightarrow q \bar{q} g$ with
$m_q=10$ GeV).
\par
There is an ongoing discussion on which is the best algorithm for
jet clustering in the case of massless quarks. The main criteria
followed to choose them are based in two requirements:
\begin{enumerate}
\item Minimize higher order corrections.
\item Keep the equivalence between parton and hadronized cross
sections.
\end{enumerate}
To our knowledge no complete comparative study of the jet-clustering algorithms
has been done for the case of massive quarks. The properties of the
different algorithms with respect to the above criteria
can be quite different in the case of massive quarks from those in
the massless case.
The first one because the leading terms containing
double-logarithms of y-cut  ($\log^2(\yc)$) that
appear in the massless calculation (at order $\as$)
and somehow determine the size
of higher order corrections are substituted in the case of massive
quarks by single-logarithms of $\yc$ times a logarithm of the quark mass.
The second one because hadronization corrections for massive quarks
could be different from the ones for massless quarks.
\par
Therefore, we will not stick to any
particular algorithm but rather present results and compare them
for all the four algorithms listed in the table~\ref{table1}.

\subsection{The  analytic calculation for the EM scheme}

Here we calculate analytically, at leading order, the three-jet decay rate
of the $Z$-boson by using the EM clustering algorithm.

At the parton level the two-jet region in the decay
$\zbbg$ is given, in terms of the
variables $y_1$ and $y_2$, by the following conditions:
\begin{equation}
\label{cuts}
y_1 < \yc\bla \mrm{or}\bla
y_2 < \yc\bla \mrm{or}\bla 1-2\rb-y_1-y_2 < \yc~.
\end{equation}
This region contains the IR singularity, $y_1=y_2=0$ and the
rate obtained by the integration of the amplitude over this part
of the phase space
should be  added to the one-loop
corrected decay width for $\zbb$. The sum of these two
quantities is of course IR finite and it is
the so-called two-jet decay width
at order $\as$.  The integration over the rest of the phase space defines
the three-jet decay width at the leading order.
It is obvious that the sum
of the two-jet and three-jet decay widths is independent of the resolution
parameter $\yc$, IR finite
and given by the quantity
$\Gamma_b = \Gamma(Z\rightarrow b\bar{b}+b\bar{b} g+\cdots)$
calculated in section~2. Therefore we have
\[
\Gamma_{b}= \Gamma^b_{2j}(\yc)+
\Gamma^b_{3j}(\yc)+\cdots~.
\]
Clearly, at order $\as$, knowing
$\Gamma_{b}$ and $\Gamma^b_{3j}(\yc)$ we can obtain
$\Gamma^b_{2j}(\yc)$ as well.
\par
The calculation of
$\Gamma^b_{3j}(\yc)$ at order $\as$ is a tree-level calculation
and does not
have any IR problem since the soft gluon region has been excluded
from phase space. Therefore the calculation can be done in
four dimensions without trouble.
\par
We will start with equation~\rfn{gammabbg} taking $\epsilon=0$
and with the phase space constrained
by the cuts defined in \eq{cuts}.
\begin{equation}
\label{gammabbgh}
\Gamma^b_{3j} = \left(C_b C_g C_F
\int d y_1 d y_2 \theta_{PS} \theta_c A_{bg}
\right)_{\epsilon =0}~,
\label{g3jdif}
\end{equation}
where the $\theta$ function
\begin{equation}
\theta_{PS} = \theta(h_p)
\end{equation}
gives the whole phase space, and the product of $\theta$ functions
\begin{equation}
\theta_c = \theta(y_2-\yc) \theta(y_1-\yc) \theta(1-2\rb-y_1-y_2-\yc)
\end{equation}
introduces the appropriate cuts for the EM  scheme.
The square of the amplitude, $A_{bg}$, is given in \eq{abg}.
The phase space and the cuts are represented in the first plot of
fig.~\ref{phase}.
\par
Depending on the value of $\yc$ the limits of integration are different,
there are three cases which correspond to three different topologies
of the overlapping of the phase space and the area defined by the cuts:
\bea
 & \yc &~<~2\rb \nonumber \\
  2\rb < & \yc &~<\bar{y}_c \nonumber \\
  \bar{y}_c < & \yc &~~~~~~~,
\eea
where $\bar{y}_c=\sqrt{\rb}(1-\sqrt{\rb})+O(\rb^2 \sqrt{\rb})$
is given by a solution of the following equation
\beq
4 (1-2 \yc -2 \rb)^2 (\yc^2+4\rb)=\yc^2(2-\yc-8\rb)^2.
\eeq

Since the
integrant is symmetric under the exchange $y_1 \leftrightarrow y_2$
we can restrict the region of integration to the region $y_1 > y_2$
(multiplying the result by a factor 2).
In addition it is useful to change variables as before,
$y_2 = z y_1$. We will not discuss the technical details of the calculation
here;
all of the integrals can be reduced to logarithmic and
dilogarithmic functions and
the final result can be written in the following form
\begin{equation}
\label{gamma3jets0}
\Gamma^b_{3j} = C_b C_g
\left( g_V^2 H\el{0}_V(\yc,\rb)+g_A^2 H\el{0}_A(\yc,\rb)\right)~,
\label{g3j}
\end{equation}
where
the superscript $\el{0}$ in the functions $H\el{0}_{V(A)}(\yc,\rb)$
reminds us that this is only the lowest order result.
Analytical expressions for the functions
$H\el{0}_V(\yc,\rb)$ and $H\el{0}_A(\yc,\rb)$
are given in appendix~C.
Obviously, the general form \rfn{g3j} is independent of
what particular jet-clustering algorithm has been used.
\par
In the limit of zero masses, $\rb=0$, chirality is conserved and the
two functions $H\el{0}_V(\yc,\rb)$
and $H\el{0}_A(\yc,\rb)$ become identical
\[
H\el{0}_V(\yc,0) =  H\el{0}_A(\yc,0) \equiv A\el{0}(\yc)~.
\]
In this case we obtain the known result for the JADE-type
algorithms, which
is expressed in terms of the function $A\el{0}(\yc)$
also given in
appendix~C \footnote{Note that with our normalization
$A^{(0)}(\yc) = \frac{1}{2} A(\yc)$,
with $A(\yc)$ defined in ref.~\cite{bethke}.}.

To see more clearly the size of mass effects we are going to study
the following ratio of jet fractions
\begin{equation}
\label{r30}
R^{bd}_3 \equiv \frac{\Gamma^b_{3j}(\yc)/\Gamma^b}
{\Gamma^d_{3j}(\yc)/\Gamma^d}=
\left(c_V \frac{H\el{0}_V(\yc,\rb)}{A\el{0}(\yc)} +
c_A \frac{H\el{0}_A(\yc,\rb)}{A\el{0}(\yc)}\right)
\left(1+6 \rb c_A + O(\rb^2)\right)~,
\end{equation}
where we have defined
\[
c_V= \frac{g_V^2}{g_V^2+g_A^2}~,\bla
c_A= \frac{g_A^2}{g_V^2+g_A^2}~.
\]
In \eq{r30} we have kept only the lowest order terms in $\as$ and
$\rb$.  The last factor is due to the normalization to total rates.
This normalization is important from the experimental point of view
but also from the theoretical point of view because in these quantities
large weak corrections dependent on the top quark mass \cite{mybb}
cancel. Note that, for massless quarks, the ratio
$\Gamma^d_{3j}(\yc)/\Gamma^d$ is independent on the neutral current
couplings of the quarks and, therefore,
 it is the same for up- and down-quarks
and given by the
function $A\el{0}$. This means that we could equally use the normalization
to any other light quark or to the sum of all of them (including also
the c-quark if its mass can be neglected).

\subsection{Estimate of higher order contributions}

All previous results come from a
tree-level calculation, however,
as commented in the introduction,
we do not know what is the
value of the mass we should use in the final results  since the
difference among the pole mass, the running mass at $\mu=m_b$ or
the running mass at $\mu= m_Z$ are next-order effects in $\as$.
\par
In the case of the
inclusive decay rate we have shown that one
could account (with very good precision)
for higher order corrections by using the running
mass at the $m_Z$ scale in the lowest order calculations.
Numerically the effect of running the quark mass from $m_b$ to $m_Z$
is very important.
\par
One could also follow a similar
approach  in the case of jet rates and try to account
for the next-order corrections by using the running quark mass at
different scales.
We will see below that the dependence of $R^{bd}_3$ on the quark
mass is quite strong (for all clustering schemes);
using the different masses (e.g. $m_b$ or $\bar{m}_b(M_Z)$)
could amount to almost a factor 2 in the mass effect.
This suggests that
higher order corrections could be important.
Here, however, the situation is quite different, since in the
decay rates to jets we have an additional scale given
by $\yc$, $E_c \equiv m_Z \sqrt{\yc}$, e.g. for $\yc=0.01$ we have
$E_c=9$~GeV and for $\yc=0.05$, $E_c=20$~GeV.
Perhaps one can absorb large logarithms, $\log(m_b/m_Z)$
by using the running coupling and the running mass
at the $\mu=m_Z$ scale,
but there will remain logarithms of the resolution parameter,
$\log(\yc)$. For not very small $\yc$ one can expect that the
tree-level results obtained by using
the running mass at the $m_Z$ scale are a good approximation,
however, as we already said,
the situation cannot be settled completely until
a next-to-leading calculation including mass effects is available.
\par
Another way to estimate higher order effects in
$R^{bd}_3$ is to use the known results for the massless case
\cite{ellis,bethke,yb}
\par
Including  higher order corrections
the general form of \eq{gamma3jets0} is still valid
with the change
$H_{V(A)}\el{0}(\yc,\rb) \rightarrow H_{V(A)}(\yc,\rb)$. Now we can
expand the functions $H_{V(A)}(\yc,\rb)$ in $\as$ and factorize the
leading dependence on the quark mass
as follows
\begin{equation}
H_{V(A)}(\yc,\rb) = A\el{0}(\yc)+\ap A\el{1}(\yc)+
\rb\left( B_{V(A)}\el{0}(\yc,\rb)
+\ap B_{V(A)}\el{1}(\yc,\rb)\right) + \cdots~.
\end{equation}
In this equation we already took into account that for massless
quarks vector and axial contributions are
identical\footnote{This is not
completely true at $O(\as^2)$
because the triangle  anomaly: there are one-loop
triangle diagrams
contributing to $Z\rightarrow b\bar{b} g$  with the top and the bottom quarks
running in the loop. Since $m_t \not= m_b$ the anomaly cancellation is
not complete. These diagrams contribute to the axial part even for
$m_b=0$ and  lead to a deviation from $A\el{1}_V(\yc) = A\el{1}_A(\yc)$
\cite{hagiwara}.
This deviation is, however, small \cite{hagiwara} and we are
not going to consider its effect here.}
\par
Then, we can rewrite
the ratio $R^{bd}_3$, at order $\as$, as follows
\bea
\label{r3a}
R^{bd}_3 =
1+ \rb\left[
 c_V
\frac{B_V\el{0}(\yc,\rb)}{A\el{0}(\yc)}\left(
1+\ap \left(\frac{B_V\el{1}(\yc,\rb)}{B_V\el{0}(\yc,\rb)}
-\frac{A\el{1}(\yc)}{A\el{0}(\yc)}
\right)\right)\right. \nonumber \\
\left. +
  c_A
\frac{B_A\el{0}(\yc,\rb)}{A\el{0}(\yc)}\left(
1+\ap\left(\frac{B_A\el{1}(\yc,\rb)}{B_A\el{0}(\yc,\rb)}
- \frac{A\el{1}(\yc)}{A\el{0}(\yc)}
\right)\right)
\right] \nonumber \\
\times\left(1+ 6\rb\left(c_A (1+2 \ap\log(\rb ))-c_V 2 \ap\right)\right)~.
\eea
 From the calculations in this paper we know
$B_V\el{0}(\yc,\rb)$  and $B_A\el{0}(\yc,\rb)$; the lowest order
function for the massless case, $A\el{0}(\yc)$,
is also known analytically for JADE-type
algorithms, \eq{massless} and refs.~\cite{bethke,yb}, and for the
DURHAM algorithm \cite{durham}.
A parameterization of the  function $A\el{1}(\yc)$ can be found
in \cite{bethke} for the different algorithms\footnote{With our
choice of the normalization $A\el{1}(\yc)=B(\yc)/4$, where $B(\yc)$
is defined in \cite{bethke}.}.
As we already mentioned
this function is different for different clustering algorithms.
The only
unknown functions in \eq{r3a}
are $B_V\el{1}(\yc,\rb)$  and $B_A\el{1}(\yc,\rb)$,
which
must be obtained from  a complete calculation at order $\as^2$ including
mass effects (at least at leading order in $\rb $).
\par
Nevertheless, in order to estimate
the impact of higher order corrections in our calculation
we will assume that
$B_{V,A}\el{1}(\yc,\rb)/B_{V,A}\el{0}(\yc,\rb) \ll
A\el{1}(\yc)/A\el{0}(\yc)$
and take
$A\el{1}(\yc)/A\el{0}(\yc)$ from\footnote{For the
EM algorithm this function has not yet been computed. To make
an estimate of higher order corrections we will use in this case
the results for the E algorithm.} \cite{bethke,yb}.
Of course this does not need to be the case but at least it gives
an idea of the size of higher order corrections.
We will illustrate the numerical effect of these corrections for
$R^{bd}_3$ in the next subsection.
As we will see, the estimated effect of next-order corrections is quite
large, therefore in order to obtain the $b$-quark mass
from these ratios the calculation of the
functions $B_{V,A}\el{1}(\yc,\rb)$ is mandatory \cite{ournext}.

\subsection{Numerical results for $R^{bd}_3$ for different clustering
algorithms}

To complete this section we present
the numerical results for $R^{bd}_3$ calculated with the different
jet-clustering algorithms.
For the JADE, E and Durham algorithms we obtained the three-jet rate
by a numerical integration over the phase-space given by the cuts
(see fig.~\ref{phase}).
For the EM scheme we used our analytical results which were also
employed to cross check  the numerical procedure.
\par
In fig.~\ref{r3} we present the ratio $R^{bd}_3$, obtained by
using the tree-level expression, \eq{r30}, against $\yc$
for $m_b= 5$~GeV and $m_b= 3$~GeV. We also plot the results given by
\eq{r3a} (with $B_{V,A}\el{1}(\yc,\rb)/B_{V,A}\el{0}(\yc,\rb)=0$)
for $m_b=5$~GeV, which gives an estimate of higher
order corrections.
For $\yc < 0.01$ we do not
expect the perturbative calculation to be valid.

As we see from the figure, the behaviour of $R^{bd}_3$ is quite different in
the different schemes. The mass effect has a negative sign for all schemes
except for the  E-algorithm.
For $\yc > 0.05$
the mass effects are at the 4\% level for $m_b= 5$~GeV and
at the 2\% level for $m_b= 3$~GeV (when the tree level expression is used).
Our estimate of higher order effects, with
the inclusion of the next-order effects in $\as$
for massless quarks, shifts the curve for $m_b=5$~GeV in the direction
of the 3~GeV result and amounts to about of  20\% to 40\% of the difference
between the tree-level calculations with the two different masses.
For both E and EM schemes we used the higher order results
for the E scheme.

For the JADE algorithm we have also plotted in fig.~\ref{r3}
the experimental results for $R^{bd}_3$
obtained by the DELPHI group \cite{delphi} on the basis
of the data collected in 1990-1991.
The experimental errors, due to the limited statistics analyzed,
are rather large.
However, one can already see the effect of the quark mass.
If the $b$-quark mass would be zero,
one should obtain a ratio $R^{bd}_3$ constant and equal to 1.
It is clearly seen from the figure that
for $\yc < 0.08$ the
data are significantly below 1.
For larger values of $\yc$, the number
of events decreases, the errors
become too large and the data are consistent with 1.
When larger amount of data is analyzed and the experimental
error is decreased, it will be very interesting to see if
data will exhibit the
different signs of the mass
effect in $R^{bd}_3$ (positive for the E scheme and negative for
the other schemes) as
predicted by our parton level calculations
(see fig.~\ref{r3}).

In spite of the fact that the effect of the quark mass in $R^{bd}_3$
has been seen, it is too early, in our opinion, to extract now
the value of the $b$-quark mass from the data.
As discussed  above
the higher order corrections
to $R^{bd}_3$ are presumably rather large and should be included in
the theoretical calculations.
However, it is clear, that once
the essential next-to-leading order corrections will be available
and all LEP data will be included in the analysis,
the ratios $R^{bd}_3$ will certainly
allow for a reasonable determination of the $b$-quark
mass and for a check of its running from $m_b$ to $m_Z$.
\par
To simplify the use of our results we present
simple fits to the ratios
$B_{V,A}\el{0}(\yc,\rb)/A\el{0}(\yc)$, which define $R^{bd}_3$ at
lowest order, for the different clustering algorithms.
We use the following parameterization:
\beq
\label{parfit}
B_{V,A}\el{0}(\yc,\rb)/A\el{0}(\yc)=
\sum_{n=0}^{2} k_{V,A}\el{n} \log^n{\yc}~,
\eeq
and the results of the fits for the range $0.01 < \yc <0.2$
are presented in table~\ref{table2}.
\begin{table}
\begin{center}
\begin{tabular}{||l|ccc|ccc||}
\hline
Algorithm & $k_V\el{0}$ & $k_V\el{1}$ & $k_V\el{2}$ &
$k_A\el{0}$ & $k_A\el{1}$ & $k_A\el{2}$ \\
\hline\hline
EM & -2.72  & -14.64 & -28.58 & -2.61  & -13.54 & -30.67 \\
\hline
JADE  & -2.01  & - 5.19 & -13.25 & -1.90  & -4.13  & -15.42 \\
\hline
E     &  4.68  &  19.04 &  25.97 &  4.71  &  19.81 &  23.39 \\
\hline
DURHAM& -1.69  & - 4.76 & -12.70 & -1.65  & -4.28  & -15.48 \\
\hline
\end{tabular}
\end{center}
\caption{Results of the tree parameter fits of the functions
$B_{V,A}\el{0}(\yc,\rb)/A\el{0}(\yc)
=\sum_{n=0}^{2} k_{V,A}^{(n)} \log^n{\yc}$ in the range $0.01 <\yc<0.2$}
\label{table2}
\end{table}

In fig.~\ref{bes} we plot the ratios
$B_{V,A}\el{0}(\yc,\rb)/A\el{0}(\yc)$ as a function of $\yc$ for
the different
algorithms (dashed lines for $m_b=5$~GeV, dotted lines for $m_b=3$~GeV
and solid curves for the result of our fits).
As we see from the figure the remnant mass
dependence in these ratios (in the range of masses we are interested in
and in the range of $\yc$ we have considered)
is rather small and for actual fits we used the average of the ratios
for the two different masses.
We see from these figures that such a simple three-parameter fit works
reasonably well for all the algorithms.

Concluding this section we would like to make the following remark.
In this paper we discuss the
$Z$-boson decay. In LEP experiments one studies the process
$e^+e^- \rightarrow (Z \gamma^*)\rightarrow b\bar{b}$ and, apart
from the resonant
$Z$-exchange cross section, there are contributions from  the pure
$\gamma$-exchange and from the $\gamma-Z$-interference.
The non-resonant
$\gamma$-exchange contribution
at the peak is less than 1\% for muon production
and in the case
of $b$-quark production there is an additional suppression factor
$Q_b^2=1/9$.
In the vicinity of the
$Z$-peak the interference
is also suppressed because it is proportional  to $Q_b (s-m_Z^2)$ ($\sqrt s$
is the $e^+e^-$ centre of mass energy). We will neglect these terms as they
give negligible contributions compared with the uncertainties in higher
order QCD corrections to the quantities we are considering.
\par
Obviously, QED initial-state radiation should be taken into account
in the real analysis;
the cross section for $b$-pair production
at the $Z$ resonance can be written as
\beq
\sigma_{b\bar{b}}(s)=\int \sigma^0_{b\bar{b}}(s') F(s'/s)ds'
\eeq
where $F(s'/s)$ is the well-known QED radiator for the total cross
section \cite{radiator} and, the Born cross section, neglecting pure $\gamma$
exchange contribution and the $\gamma-Z$-interference, has the form
\beq
\sigma^0_{b\bar{b}}(s)=\frac{12 \pi \Gamma_e \Gamma_b}{m_Z^2}
\frac{s}{(s-m_Z^2)^2+m_Z^2\Gamma_Z^2}
\eeq
with obvious notation. Note that $\Gamma_b$ in this expression
can be an inclusive width as well as some more exclusive quantity,
which takes into account some kinematical restrictions on the final state.

\mysection{Angular distribution}

If the gluon jet can be identified with enough efficiency, an
interesting quantity which is very sensitive to the IR behaviour of the
amplitudes is the angular
distribution with respect to the angle formed between one of the
quark jets and the gluon jet\footnote{We thank J. Fuster for
suggesting us the study of this observable.}.
If $\vartheta_1$ ($\vartheta_2$) is the
angle between the quark (antiquark) and the gluon jets we define
$\vartheta=\min(\vartheta_1,\vartheta_2)$. We want to obtain the angular
distribution with respect to $\vartheta$. The starting point is
\eq{gammabbgh} where we change variables from one of the
$y_1$ or the $y_2$ variables to $\vartheta$.
To do this we take into account that the amplitude is completely
symmetric in $y_1$ and $y_2$, therefore we can restrict the integration
only to the region $y_2 > y_1$ and add a factor 2. In that region
$\vartheta = \vartheta_1$. Therefore to obtain the distribution with respect
to $\vartheta$ it is enough to obtain the distribution with respect
to $\vartheta_1$ but constraining the phase space integration to
$y_2 > y_1$.

For $y_2> y_1$ we can easily express $y_1$ in terms of
$\cos \vartheta=\cos \vartheta_1$ as follows
\begin{equation}
y_1= \frac{y_2 \left(1-y_2-\cos\vartheta\sqrt{(1-y_2)^2-4\rb}\right)}
        {1+y_2+\cos\vartheta\sqrt{(1-y_2)^2-4\rb}}~.
\label{y1c1}
\end{equation}
Adding the Jacobian of the transformation we find from \eq{gammabbgh}
(taking $\epsilon=0$ as this quantity is IR convergent)
\begin{equation}
\label{angdis}
\frac{d\Gamma^b_{3j}}{d\vartheta} = C_b C_g C_F 2
\int d y_2 \theta_{PS} \theta_c\ \theta(y_2-y_1)
\sin \vartheta\frac{2 y_2 \sqrt{(1-y_2)^2-4\rb})}
{\left(1+y_2+\cos\vartheta\sqrt{(1-y_2)^2-4\rb}\right)^2}\ A_{bg}~,
\end{equation}
where $y_1$ is expressed in terms of $\cos\vartheta$ and $y_2$ using
\eq{y1c1}.

In order to see how large mass effects are in this angular distribution
we define the following ratio of angular distributions:
\begin{equation}
\label{ratangdis}
R^{bd}_\vartheta= \left.
\frac{1}{\Gamma^b}\frac{d\Gamma^b_{3j}}{d\vartheta}\right/
\frac{1}{\Gamma^d}\frac{d\Gamma^d_{3j}}{d\vartheta}
\end{equation}

In the case of massless quarks
the integration limits
in \eq{angdis}
can be found analytically for the JADE-type schemes
and the result of the integration over $y_2$
is expressed in terms of logarithms involving $\vartheta$ and
$\yc$.
We find
\begin{equation}
\label{angdis0}
\frac{1}{\Gamma^d}\frac{d\Gamma^d_{3j}}{d\vartheta} =
\frac{\alpha_s}{\pi} f_\vartheta (\yc)~,
\end{equation}
where the function $f_\vartheta(\yc)$ is given analytically in appendix~D
for the JADE-type schemes
and represented in fig.~\ref{ang0}
for the JADE-type and the Durham  algorithms for different values of
$\yc$ ($\yc = 0.02$ (solid line), $\yc = 0.04$ (dashed line)
$\yc = 0.06$ (dotted line ) and $\yc = 0.08$ (dash-dotted line)). We
observe a very sharp peak, for both algorithms, in the region
of $90^\circ$--$100^\circ$ depending on the value of $\yc$, for $\yc=0.04$
the peak is at about $92^\circ$ for the JADE-type algorithms and
at about $99^\circ$ for the Durham algorithm.
We see that the absolute size of the peak is a factor two larger
in the case of the JADE-type algorithms (for the same value of $\yc$)
than in the case of the DURHAM scheme. This is due to the difference of phase
spaces for two schemes.

For massive quarks, although the integrations can still be performed
analytically in the EM scheme,
some of the integration limits are solutions of polynomial equations of
the third degree and the analytical
result is not especially enlighting. Then,
we have computed the ratio $R^{bd}_\vartheta$ by doing the one-dimensional
integration in (\ref{angdis}) numerically.

Numerical results for $R^{bd}_\vartheta$ are presented in fig.~\ref{ang}
for the different algorithms
for $\yc=0.04$ and for both $m_b = 5$~GeV (solid line) and for
$m_b = 3$~GeV (dashed line).
In all cases we plot the ratios for the
interval of angles for which the differential cross section is still
sizable (see fig.~\ref{ang0}),
i.e. $\vartheta \approx 45^\circ-120^\circ$ for JADE-type schemes
and $\vartheta \approx 50^\circ-130^\circ $ for the DURHAM algorithm.
For small angles and $m_b=5$~GeV
the effect can be as large as 10\% of the
ratio. Note, however, that the angular distribution,
fig.~\ref{ang0}, drops down rapidly for such small angles. In
addition, since the ratio changes very fast in this region the exact
size of the effect will depend on the angular resolution achieved at
experiment.

As in the case of ratios of three-jet event rates, $R^{bd}_3$,
the variation of the ratio of angular distributions, $R^{bd}_\vartheta$,
for $m_b=5$~GeV and $m_b=3$~GeV gives a measure of the size of
higher order corrections.

We observe in all the ratios the irregular behaviour
in the region where the massless angular distribution peaks. This
is due to the fact that in the massive case the position of the peak
is slightly
shifted with respect to the massless case.
The mismatch between the two peaks appears as a discontinuity
in the ratio when seen from large scales.

It will be interesting to see if
data really follow these patterns for $R^{bd}_{\vartheta}$.
A preliminary analysis
performed by the DELPHI group\cite{joanang} seems to indicate that, indeed,
data do follow these angular distributions, at least qualitatively,
and exhibit the variations present in the different algorithms.

\mysection{Discussion and conclusions}

In this paper we have presented a theoretical
study of quark-mass effects in the decay of the
$Z$-boson into bottom quarks.

First, we have reproduced, with the complete mass dependences,
the results for the inclusive  decay rate of the
$Z \rightarrow b \bar{b}+\cdots$ to order $\as$ by
adding gluon bremsstrahlung from the $b$-quarks
to the one-loop corrected decay width of $Z\rightarrow b \bar{b}$.
Although the sum of the two contributions is finite, each of them
is separately IR divergent.
We used dimensional continuation
to regularize the IR divergences and gave a complete
analytical result in arbitrary space-time dimensions for each of the
two contributions.

The main contribution of this paper is, however, the analysis of
some three-jet observables which are more sensitive to the value of
the quark masses.

For a slight modification of the JADE
algorithm (the EM algorithm)
we have calculated analytically the three-jet decay width of the
$Z$-boson into $b$-quarks as a function of the
jet resolution parameter, $\yc$, and the $b$-quark mass.
The answer is rather involved, but can be  expressed in terms of elementary
functions.
Apart from the fact that these analytical calculations are interesting by
themselves, they can also be used to test Monte Carlo simulations.
For the EM, JADE, E and DURHAM clustering algorithms we have obtained
the three-jet decay width by a simple two-dimensional numerical integration.
Numerical and analytical results have been compared in the case of the
EM scheme.
\par
We discussed quark-mass effects by considering the quantity
\[
R^{bd}_3=\frac{\Gamma^b_{3j}(\yc)/\Gamma^b}{\Gamma^d_{3j}(\yc)/\Gamma^d}
=1 + \frac{m_b^2}{m_Z^2} F(m_b,\yc)~
\]
which has many advantages from both the theoretical and  the experimental
point of views. In particular, at lowest order,
the function $F(m_b,\yc)$ is almost
independent on the quark mass (for the small values of the mass in which
we are interested in) and has absolute values ranging from 10 to 35
(depending on $\yc$ and on the algorithm), where the larger values are
obtained for $\yc$ of about $0.01$.

At the lowest order in $\as$ we do not know what is the exact value of the
quark mass that should be
used in the above equation since the difference between the different
definitions of the $b$-quark mass,
the pole mass, $m_b \approx 5$~GeV, or the running mass
at the $m_Z$-scale, $\bar{m}_b(m_Z) \approx 3$~GeV, is order $\as$. Therefore,
we have presented all results for these two values of the mass and
have interpreted
the difference as an estimate of higher order corrections. Conversely one
can keep the mass fixed and include in $F(m_b,\yc)$ higher order corrections
already known for the massless case.
According to these estimates
the $O(\as)$ corrections can be about 40\% of the tree-level
mass effect (depending on the clustering scheme),
although we cannot exclude even larger corrections.

By using the lowest order result we find that for moderate
values of the resolution parameter, $\yc \approx 0.05$, the mass
effect in the ratio $R^{bd}_3$ is about $4\%$
if the pole mass value of the $b$-quark, $m_b \approx 5$~GeV, is used,
and the effect decreases to 2\% if $m_b=3$~GeV.

We have compared our predictions for $R^{bd}_3$ for the
JADE algorithm, with the results obtained from the
1990-1991 data
by the DELPHI group \cite{delphi}. Although the
errors obtained in the analysis of this limited sample of
data are rather large, especially for $\yc > 0.08$,
one clearly sees that for small values of y-cut
($\yc < 0.08$) the experimental points are systematically below 1,
thus clearly exhibiting
the effect of the mass of the quark, as for massless quark $R^{bd}_3=1$.
The size of the effect agrees
roughly with the predictions.
One can expect the reduction of the experimental error by, at least,
a factor two when the data collected in 1992 are included in the analysis.
Then, mass effects will be more clearly seen
and it will be very interesting to see
if data follow the different qualitative behaviour of
the ratio $R^{bd}_3$ as a function of $\yc$ as predicted by
the parton model calculations (positive effect for the E scheme and
negative mass effect for the other algorithms).
However, in order to extract a meaningful value of the $b$-quark mass
from the data it will be necessary to include next-to-leading order
corrections since the leading mass effect we have calculated does
not distinguish among the different definitions of the quark mass
(pole mass, running mass at the $m_b$ scale or running
mass at the $m_Z$ scale).
We believe that
the future analysis of the whole
LEP statistics and its comparison with the theoretical predictions
for the three-jet ratios, which meet the future experimental precision,
will allow for a good determination of the $b$-quark
mass at the highest energy scale
and for a check of its running from $m_b$ to $m_Z$.

The high precision achieved at LEP allows for a good separation of the
gluonic and quark jets and a measurement of the angular distribution
of the radiated gluon with respect to the quark momenta.
This angular distribution has been calculated for massless quarks
analytically for
the JADE-type schemes and numerically for the DURHAM algorithm.
We have studied the
mass effects, for the different jet-clustering algorithms, in the quantity
\[
R^{bd}_\vartheta=
\left.\frac{1}{\Gamma^b}\frac{d\Gamma^b_{3j}}{d\vartheta}\right/
\frac{1}{\Gamma^d}\frac{d\Gamma^d_{3j}}{d\vartheta}~.
\]
We have shown that,
for a reasonable value of the resolution parameter, $\yc = 0.04$,
the mass effects in this ratio can be as large as 10\% of the ratio
for $m_b=5$~GeV (depending
on the algorithm, the angle $\vartheta$ and the angular resolution).
The larger values
are obtained for small angles where, however, the angular distribution
falls down very rapidly. A fit to this ratio can be used to extract
the value of the $b$-quark mass.
It will be interesting to see if data really follow the predictions
for the  angular
distributions and if the mass effects in the ratio of angular
distributions are well described by our results.

Concluding, we have raised the question of the possibility of measuring
the $b$-quark mass at LEP by using three-jet observables.
In our opinion, this is a big challenge
for both experimentalists and theorists.
Clearly, more work has to be done
in order the precision of theoretical predictions meet the
experimental accuracy, in particular
order $\as^2$ calculations and studies of hadronization corrections
including mass effects will be needed.
However, this effort is worth since it will allow for an independent
measurement of $m_b$ at much larger energies where, presumably,
non-perturbative effects are negligible.

\section*{Acknowledgements}

We would like to thank J.~Fuster for his continuous encouragement,
for many helpful discussions and for carefully reading the manuscript.
G. Rodrigo acknowledges the CERN theory group
for its hospitality during the preparation of this work
and the Conselleria de Cultura, Educaci\'o
i Ci\`encia de la Generalitat Valenciana for financial support.
This work was  supported in part by
CICYT, Spain, under grant AEN93-0234.

\vfill\eject

\appendix

\begin{center}{\Large\bf APPENDICES}\end{center}

\mysection{Phase space in $D=4-\epsilon$ dimensions}

The phase space for $n$-particles in the  final state
in $D$-dimensions \cite{dimreg}
$(D=4-2\epsilon)$
has the following general form
\bea
d(PS_n)
&=& (2\pi)^D \prod_{i=1,n} \frac{d^{D-1}p_i}{(2\pi)^{D-1}2E_i}
\delta^D \left( q-\sum_{i=1,n}p_i \right) \\
&=&
(2\pi)^D \prod_{i=1,n} \frac{d^{D}p_i}{(2\pi)^{D-1}2E_i}
\delta(p_i^2-m_i^2)\Theta(E_i) \delta^D
\left( q-\sum_{i=1,n}p_i \right)~.
\label{phased}
\eea
Then doing several trivial integrations we have
the following phase-space factor for the process
$Z\rightarrow b\bar{b}$
\begin{equation}
PS_2=\frac{1}{4\pi} \frac{\beta}{2}
\frac{\Gamma(1-\epsilon)}{\Gamma(2-2\epsilon)}
{\left( \frac{\beta^2 m_Z^2}{4\pi} \right) }^{-\epsilon}~,
\end{equation}
where
$\beta=\sqrt{1-4 \rb}$   with $\rb=m_b^2/m_Z^2$,

For the case of the decay into three particles,
$Z\rightarrow b\bar{b}g$, we have
\bea
d(PS_3)=\frac{m_Z^2}{16(2\pi)^3}
\frac{1}{\Gamma(2-2\epsilon)}
{\left( \frac{m_Z^2}{4\pi} \right) }^{-2\epsilon}
\theta(h_p) h_p^{-\epsilon} dy_1 dy_2~,
\eea
where the function $h_p$ which gives a phase-space boundary
in terms of variables $y_1=2(p_1 k)/m_Z^2$ and $y_2=2(p_2 k)/m_Z^2$
has the form
\begin{equation}
h_p = y_1 y_2 (1-y_1-y_2) - \rb (y_1+y_2)^2~.
\end{equation}

\mysection{Inclusive decay rate functions}

\renewcommand{\b}{\beta}
\renewcommand{\c}{(c)}

In this section we collect the functions needed in section~2. The
relevant form factors are:
\begin{eqnarray}
f_\epsilon &= & C_F
\left(1+\frac{1+\b^2}{2\b}\log\c\right)~,\\
f_{Tf} &=& C_F \frac{1-\b^2}{2\b}\log\c~,\\
f_{Af} &= & f_\epsilon \log(r_b)+
C_F\left[-2-\frac{2+\b^2}{2\b}\log\c\right.\nonumber\\
&+& \left.
\frac{1+\b^2}{\b}\left(\mrm{Li}_2\c+\frac{\pi^2}{3}-\frac{1}{4}\log^2\c
+\log\c\log\left(1-c\right)\right)\right]~.
\end{eqnarray}
In the expression for $f_{Af}$, the first term, proportional to
$\log(r_b)$, comes because
our election for the term proportional to the divergence. The
vector form factor, $f_{Vf}$ can be written in terms of the
other two form factors,
\begin{equation}
f_{Vf}=f_{Af}+f_{Tf}~.
\end{equation}
In terms of these form factors the functions $F_V$ and $F_A$ that
appear in \eq{gammab1} are
\begin{eqnarray}
F_V &=& \beta\left(\frac{(3-\beta^2)}{2}
f_{Vf}+\frac{3}{2} f_{Tf}\right)~, \\
F_A &=& \beta^3  f_{Af}~.
\end{eqnarray}

The functions that come from real bremsstrahlung can be written as
follows,
\begin{eqnarray}
G_V &=& \b\left(\half (3-\b^2) G_P+G_{Vh}\right)~,\\
G_A &=& \b\left(\b^2 G_P+G_{Ah}\right)~,
\end{eqnarray}
where
\begin{equation}
G_P = G_{Ph}+2f_\epsilon (1+\log(\b))~.
\end{equation}
The terms proportional to $f_\epsilon$ come again from our choice of
the coefficient of the divergence, and $G_{Ph}$ is the finite part coming
from the integration of the term proportional to $h_p$ in the amplitude
\begin{equation}
G_{Ph} = \frac{C_F}{2\b} \int_c^{1/c} dz
g(z)\frac{1+z}{z} \log\left(g(z)^3 z(1+z)\right)~.
\end{equation}
The result of the integration gives
\begin{eqnarray}
G_{Ph} &=& C_F\left[-2\log\left(\frac{4\b^3}{1-\b^2}\right)
+2-\frac{2+\b^2}{\b}\log\c \right.\nonumber\\
&&\hspace{-2cm}\left.
-\frac{1+\b^2}{\b}\left(\frac{1}{4}\log^2\c+\frac{\pi^2}{3}
-\mrm{Li}_2\c-\mrm{Li}_2(c^2)
-3\log\c\log\left(1+c\right)\right)\right]~.
\end{eqnarray}
The functions $G_{Vh}$ and $G_{Ah}$ come from the integration of the
$h_V$ and $h_A$ terms respectively
\begin{equation}
G_{Vh} = \frac{C_F}{4\b}\int_c^{1/c} dz g(z)^2 \left(z+\frac{1}{z}\right) =
-\frac{C_F}{8}\left(9+\b^2+\frac{9-2\b^2+\b^4}{2\b}\log\c\right)~,
\end{equation}
and
\begin{equation}
G_{Ah} = \half (3-\b^2) G_{Vh}+(1-\b^2) \tilde{G}_{Ah}~,
\end{equation}
where
\begin{equation}
\tilde{G}_{Ah} = \frac{C_F}{4\b} \int_c^{1/c} dz g(z)^2 =
\frac{C_F}{8}\left(3-\b^2+\frac{3-2\b^2-\b^4}{2\b}\log\c\right)~.
\end{equation}

\mysection{Three-jet event rate functions}

The functions $H^{(0)}_V$ and $H^{(0)}_A$, which give the leading
contribution to the three-jet
decay rate in the EM algorithm, can be written in the following form
\begin{eqnarray}
H^{(0)}_V(\yc,\rb) & = &
C_F\sum_{i=1,3}\theta_i
\left[\frac{(3-\be^2)}{2}K^{i}_S+K^{i}_{V}\right] \nonumber\\
H^{(0)}_A(\yc,\rb) & = &
C_F\sum_{i=1,3}\theta_i
\left[\be^2 K^{i}_S+\frac{(3-\be^2)}{2}K^{i}_V
+(1-\be^2) K^{i}_{A}\right]~.
\end{eqnarray}
with
\bea
\theta_1 & = & ~ \theta(\bar{\yc}-\yc)  \nonumber\\
\theta_2 & = & - \theta(\yc-2\rb)\theta(\bar{\yc}-\yc) \nonumber\\
\theta_3 & = & ~ \theta(\yc-\bar{\yc})\nonumber
\eea
and $\bar{\yc} \approx \sqrt{\rb}(1-\sqrt{\rb})$.
Here $K^{i}_S$ corresponds to the soft part and $K^{i}_{V(A)}$
to the vector (axial) hard part. These functions are given by
\bea
K^{1}_S & = &
4\yc (\zb^{-1}-1)-2\left(\frac{1-\be^2}{2}+2\yc\right)\log(\zb)
 + 4\be \log\left(\frac{\zb - c}{1 - \zb c}\right) \nonumber\\
& + & (1-\be^2)\left[1 + \zb^{-1} - 2\zb  +
\left(\zb^{-1} - \zb\right)
\log\left(\frac{\yc(1 + \zb)(1 + c)^2}{(\zb - c)(1 - \zb c)} \right)
\right] \\
& + & 2(1 + \be^2) \left[
   \frac{1}{2}\log(\zb)\log\left(\frac{\yc^2 (1 + c)^4}{\zb}\right)
-\frac{\pi^2}{12}
- \mrm{Li}_2(-\zb) - \mrm{Li}_2\left(\frac{c}{\zb}\right)
+ \mrm{Li}_2(\zb c)\right]
\nonumber\\
& ~ & ~~~~\nonumber\\
& ~ & ~~~~\nonumber\\
%
%
K^{1}_{V} & = &
1 + \frac{\yc^2 (1-\zb^{-2})}{2} -2(1 + \zb)^{-1}
  + \frac{(1-\be^2)(3+\be^2)}{8}(\zb-\zb^{-1}) \nonumber \\
&  + &   \frac{(1 - \be^2)^2}{32}(\zb^{-2} - \zb^2)
 -  \left(1+\frac{(1-\be^2)^2}{8}-\yc^2\right)\log(\zb) \\
& ~ & ~~~~\nonumber\\
& ~ & ~~~~\nonumber\\
%
%
K^{1}_{A} & = &
-\frac{1}{2}+ \yc^2(1- \zb^{-1})
+ \frac{(1-\be^2)^2}{16}(\zb^{-1} - \zb) + (1 + \zb)^{-1}
\nonumber \\
& + &
 \frac{(1-\be^2)(3+\be^2)}{8}\log(\zb) \\
& ~ & ~~~~\nonumber\\
& ~ & ~~~~\nonumber\\
& ~ & ~~~~\nonumber\\
%
%
K^{2}_S & = & -2(1 + \be^2- 2\yc) \log(\za)
  +   4\be \log\left(\frac{\za-c}{1-\za c}\right) \nonumber\\
& + & (1 - \be^2)(\za^{-1} - \za)
   \left[
 2+\log\left(\frac{(1 +\be^2 - 2\yc)\za(1+c)^2}
                  {2(\za-c)(1- c\za)}\right)\right] \nonumber\\
& + & 2(1 + \be^2)
 \left[ \log(\za)\log\left(\frac{(1 +\be^2 - 2\yc)(1+c)^2}{2}\right)
  - \mrm{Li}_2\left(\frac{c}{\za}\right) + \mrm{Li}_2(c \za) \right]\\
& ~ & ~~~~\nonumber\\
& ~ & ~~~~\nonumber\\
%
%
K^{2}_{V} & = &
-(1 +\be^2 - \yc) \yc\frac{(1-\za)}{(1 +\za)}
 + \frac{(1-\be^2)^2}{32}(\za^{-2} - \za^2) \nonumber\\
& + & \frac{(1-\be^2)(3+\be^2)}{8}(2-\za^{-1} + \za - 4(1 + \za)^{-1})
\nonumber\\
& - & \left(\frac{(1-\be^2)(7+\be^2)}{8}
 + \yc + \be^2\yc - \yc^2\right)\log(\za)  \\
& ~ & ~~~~\nonumber\\
& ~ & ~~~~\nonumber\\
%
%
K^{2}_{A} & = &
\frac{(1-\be^2)^2}{16}(\za^{-1} - \za)
 + \frac{(1 + \be^2- \yc)\yc}{2}\frac{(1 -\za)}{(1 +\za)} \nonumber\\
& + & \frac{(1 - \be^2)(3 + \be^2)}{8}
(-1 + 2(1 + \za)^{-1}+ \log(\za)) \\
& ~ & ~~~~\nonumber\\
& ~ & ~~~~\nonumber\\
& ~ & ~~~~\nonumber\\
%
%
K^{3}_S & = &
- \frac{(1 + \be^2)\pi^2}{6}
- \frac{(1 -\be^2 - 4\be^2\zg)}{\zg}\frac{(1-\zg)}{(1 + 2\zg)}
+ \frac{(1 + 3\be^2 - 2(1-\be^2)\zg)}{(1 + 2\zg)}\log(\zg)\nonumber\\
& + & (1 + \be^2)\left[
\log^2\left(\frac{\zg}{1 + \zg}\right)
+ 2\mrm{Li}_2\left(\frac{\zg}{1 +\zg}\right)\right] \\
& ~ & ~~~~\nonumber\\
& ~ & ~~~~\nonumber\\
%
%
K^{3}_{V}
& = & \frac{(1 +\be^2)^2}{8}
\left(-\frac{3(1-\zg^2)}{(1+2\zg)^2}-\frac{2}
{(1+2\zg)}\log(\zg)\right) \\
& ~ & ~~~~\nonumber\\
& ~ & ~~~~\nonumber\\
%
%
K^{3}_{A} & = & \frac{(1 +\be^2)^2}{8}
\left(\frac{1 -\zg}{1 + 2\zg}\right)^2~,
\eea
where we used the following notation,
\bea
\za & = & \frac{1}{2\rb}\left(\yc-\sqrt{\yc^2-4\rb^2}\right)~, \nonumber\\
\zb & = & \frac{1}{2\rb}
\left(1-\yc-2\rb-\sqrt{(1-\yc)^2-4\rb}\right)~,\nonumber\\
\zg & = & \frac{\yc}{1-2\rb-2\yc}~, \nonumber\\
c & = & \frac{1-\be}{1+\be}~.
\eea

In the limit of massless quarks, $\rb \rightarrow 0$,
from the functions $H^{(0)}_V$ and $H^{(0)}_A$ given above
we obtain
\beq
H^{(0)}_V(\yc,\rb \rightarrow0)= H^{(0)}_A(\yc,\rb \rightarrow 0)
 \rightarrow A^{(0)}~,
\eeq
Here the function $A^{(0)}(\yc)$ is the
known result \cite{bethke,yb} for the JADE algorithm
\bea
A^{(0)}(\yc) & = & 2 C_F\left[
-\frac{\pi^2}{3}+\frac{5}{2}-6\yc-\frac{9}{2}\yc^2
+(3-6\yc)\log \left( \frac{\yc}{1-2\yc} \right)\right. \nonumber \\
& + & \left. 2\log^2 \left( \frac{\yc}{1- \yc} \right)
+4\mrm{Li}_2\left(\frac{\yc}{1-\yc}\right) \right]~.
\label{massless}
\eea
The function $A(\yc)$ given in refs.~\cite{bethke,yb} differs from our
$A\el{0}(\yc)$ in a factor 2 because we chose a different normalization
for it.

\mysection{Angular distribution functions}
The angular distribution studied in section~4 is given,
in the massless case,  by the function $f_\vartheta(\yc)$. In
the JADE-type algorithms it can be written as follows
\begin{equation}
f_\vartheta(\yc) = C_F \sin(\vartheta) \sum_{i=1,2} \theta_i f_i(\yc)~,
\end{equation}
where the $theta_i$ functions have the form
\begin{eqnarray*}
\theta_1 &=& \theta\left(\cos\vartheta+\frac{\yc}{(1-\yc)}\right)
\theta\left(\frac{(1-6\yc+\yc^2)}{(1+\yc)^2}-\cos\vartheta\right)~,\\
\theta_2 &=& \theta\left(-\frac{\yc}{(1-\yc)}-\cos\vartheta\right)
\theta\left(\cos\vartheta+\frac{1-\yc}{1+\yc}\right)~,
\end{eqnarray*}
and
\begin{eqnarray}
f_1(\yc) &=&\frac{(1+b)^2}{b}\left[\frac{\yc+8b-3}{4}\sqrt{(1-\yc)^2-4b\yc}
\right.\nonumber\\
&-& \left. \log\left(\frac{x_1}{x_2}\right)+
(1+b+2b^2)\log\left(\frac{b+x_1}{b+x_2}\right)\right]~,\\
f_2(\yc) &=& \frac{(1+b)^2}{b}\left[
\frac{(\yc-b)(\yc^2+2\yc-2b\yc-5b^2)}{4(b+\yc)}\right.\nonumber\\
&-& \left. b (1+2b) \log\left(\frac{2b}{b+\yc}\right)-
\log\left(\frac{1-b}{1-\yc}\right)\right]~.
\end{eqnarray}
In these equations we defined
\begin{eqnarray*}
b &=& \frac{1+\cos\vartheta}{1-\cos\vartheta}~,\\
x_1 &=& \frac{1}{2}\left(1-\yc-\sqrt{(1-\yc)^2-4b\yc}\right)~,\\
x_2 &=& \frac{1}{2}\left(1-\yc+\sqrt{(1-\yc)^2-4b\yc}\right)~.
\end{eqnarray*}

\vfil\eject

\vfil\eject

%
%
%
\def\mafigura#1#2#3#4{
  \begin{figure}[hbtp]
\vspace{1cm}
    \caption{#3}
    \label{#4}
  \end{figure} }
\mafigura{16cm}{feynman.ps}
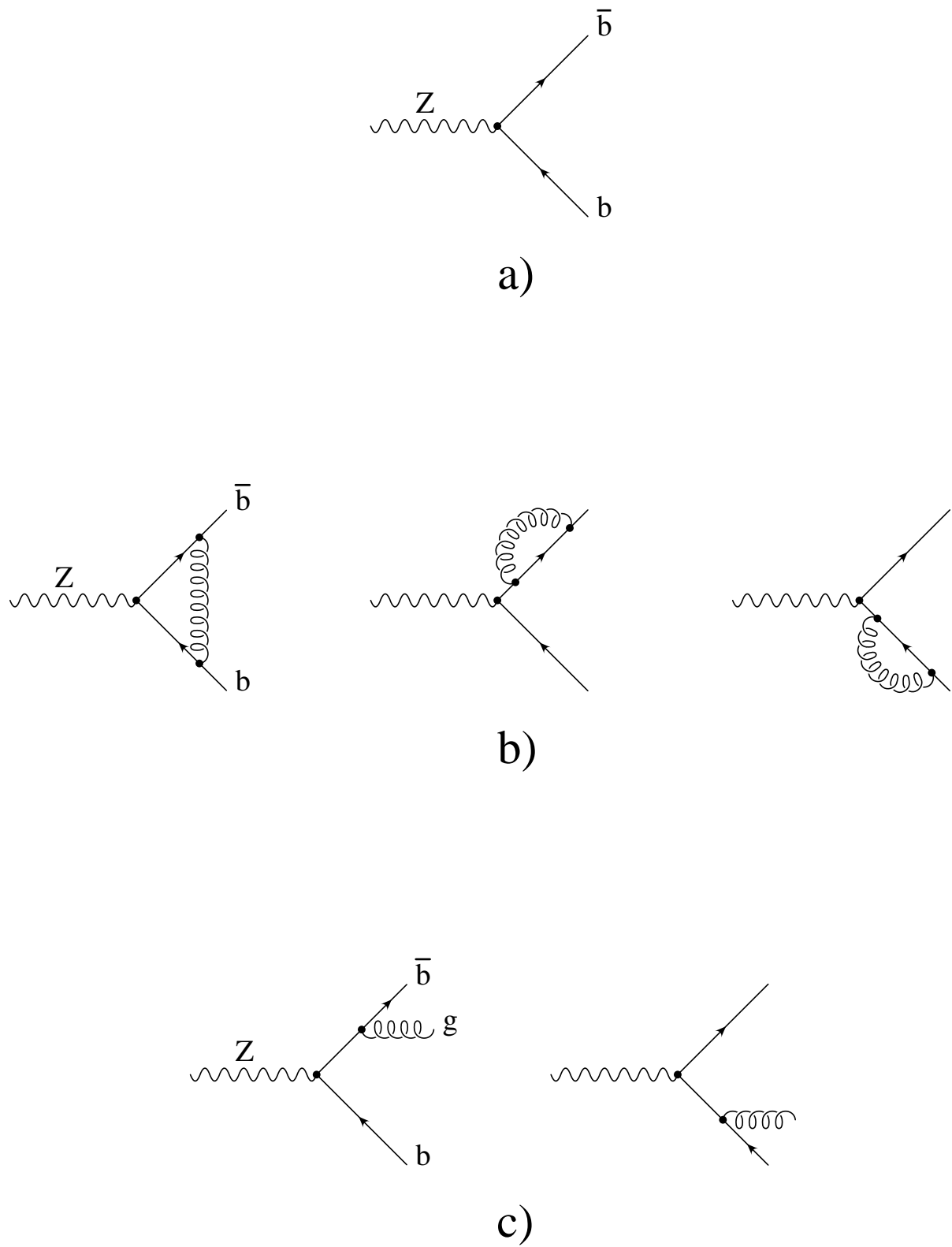
{Feynman diagrams contributing to the decay rates
$Z\rightarrow b\bar{b}$, $Z\rightarrow b\bar{b} g$
 at order $\as$.}
{feynman}
\mafigura{16cm}{phase.ps}
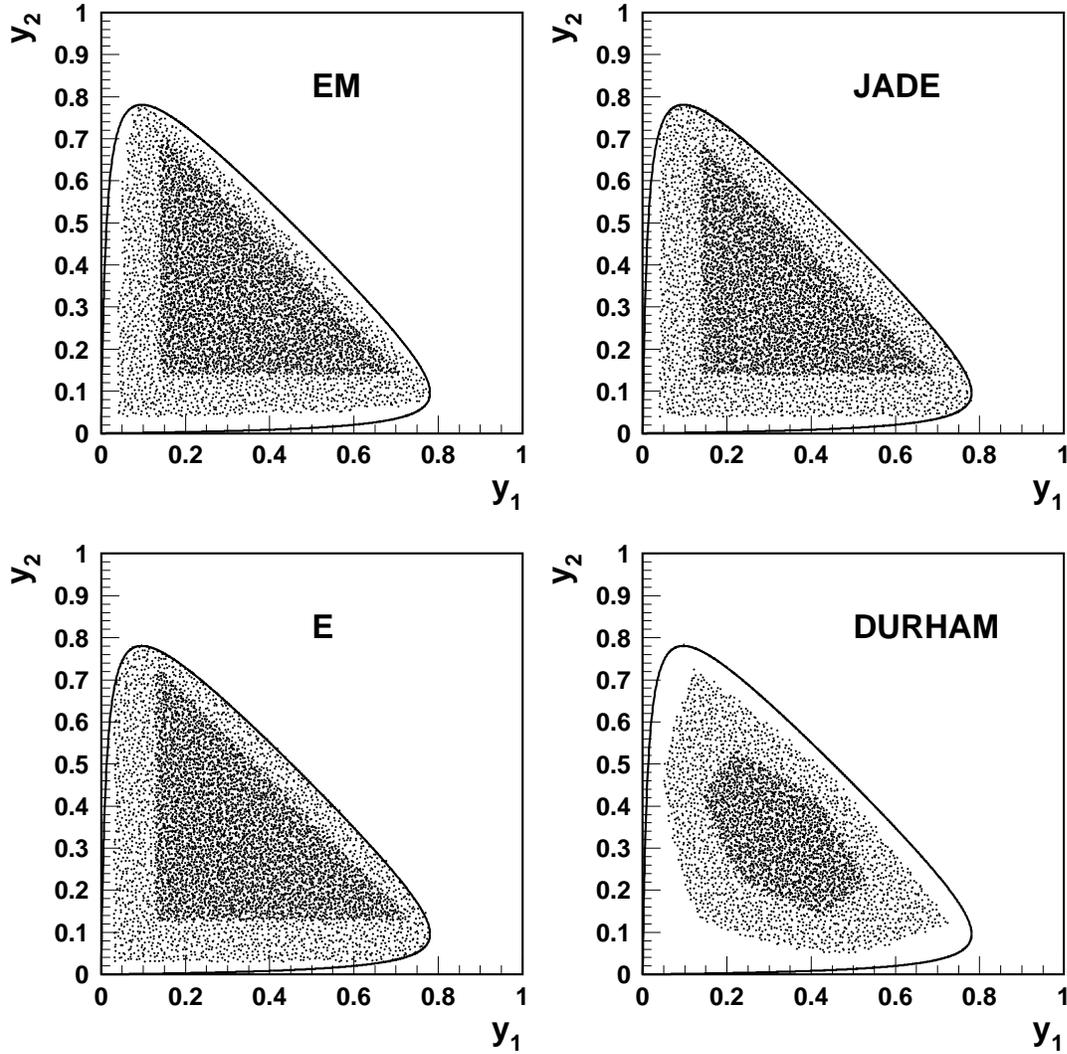
{The phase
space for $Z\rightarrow b\bar{b} g$ in the plane $y_1$ and $y_2$
with cuts ($\yc=0.04$ and $\yc=0.14$) for the different algorithms. The mass
of the quark has been set to 10 GeV to enhance mass effects in the plot.}
{phase}
\mafigura{16cm}{r3.ps}
{The ratios $R^{bd}_3$ (see \protect\eq{r30}) for the four algorithms.
Solid lines correspond to $m_b=5$~GeV,
dashed lines correspond to $m_b=3$~GeV and dotted lines give our estimate
of higher order corrections to the $m_b=5$~GeV curve. For the JADE
algorithm we have also included the results of the
analysis of the data collected during 1990-1991 by
the DELPHI group \protect\cite{delphi}.}
{r3}
\mafigura{16cm}{bes.ps}
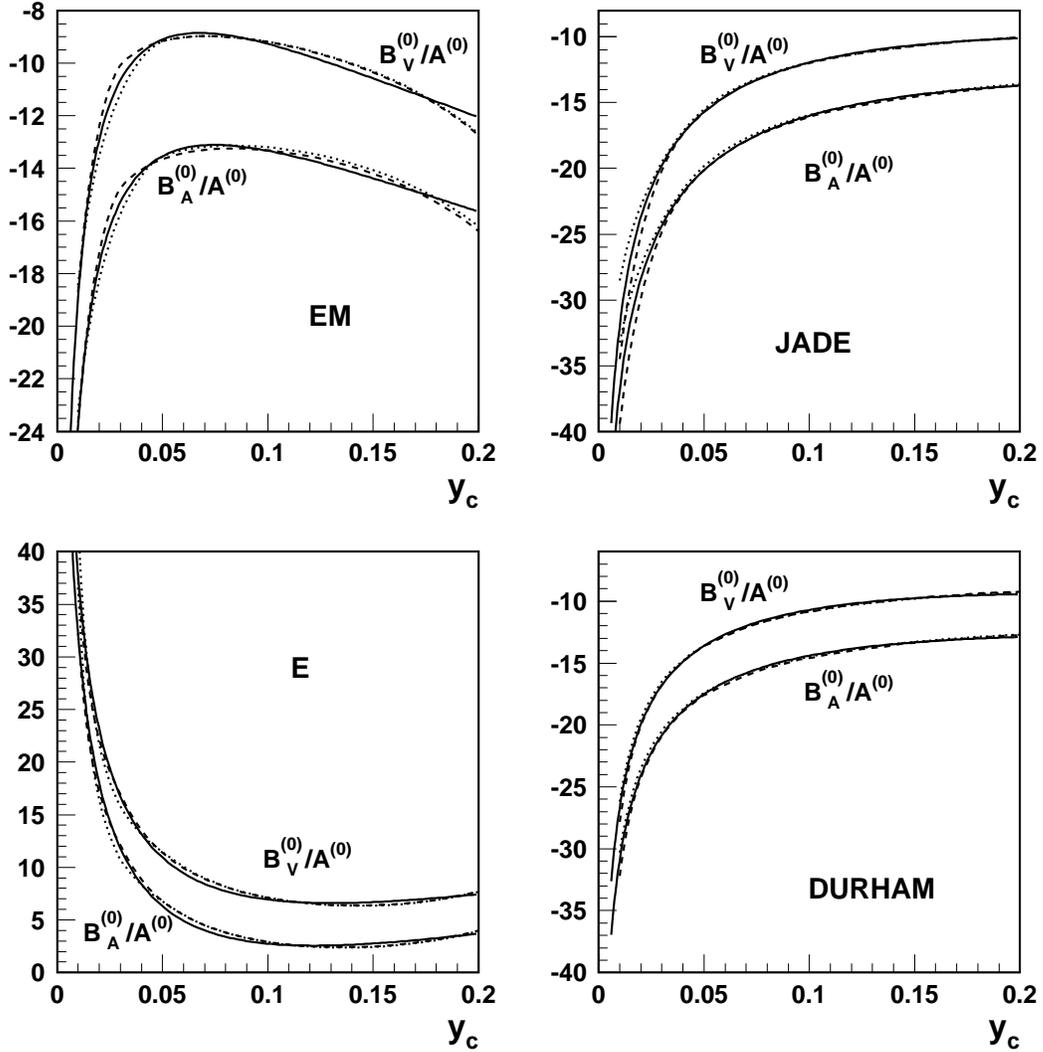
{The functions $B_V\el{0}/A\el{0}$ and $B_A\el{0}/A\el{0}$ for the four
algorithms. Dashed lines for $m_b=3$~GeV, dotted lines for $m_b=5$~GeV
and solid lines for our three-parameter fit, \protect\eq{parfit}.}
{bes}
\mafigura{16cm}{ang0.ps}
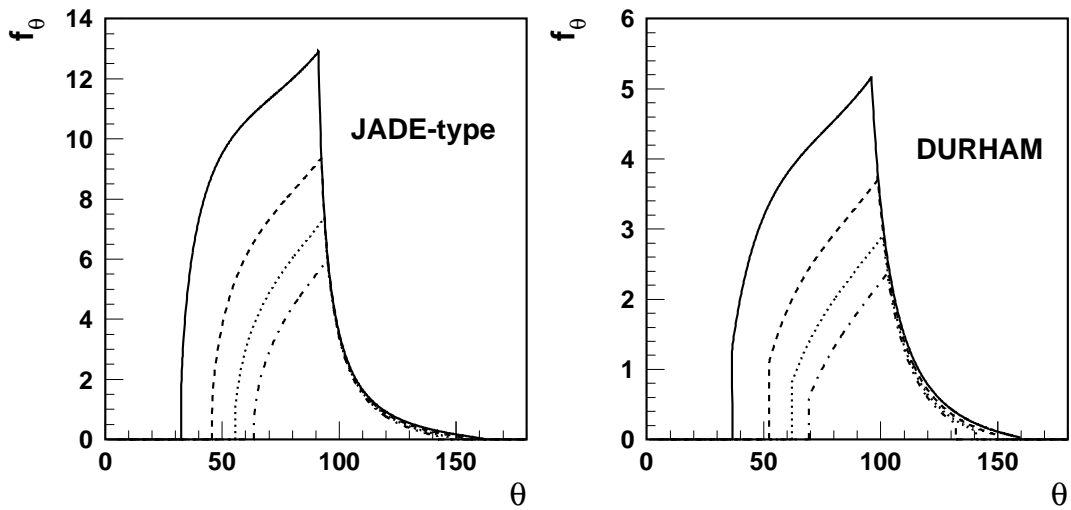
{Normalized angular distributions (\protect\eq{angdis0})
with respect to the angle formed between the quark and the gluon jets
for the massless case for JADE-type and
DURHAM algorithms. Solid line for $\yc=0.02$, dashed line for $\yc=0.04$,
dotted line for $\yc=0.06$ and dash-dotted line for $\yc=0.08$}
{ang0}
\mafigura{16cm}{ang.ps}
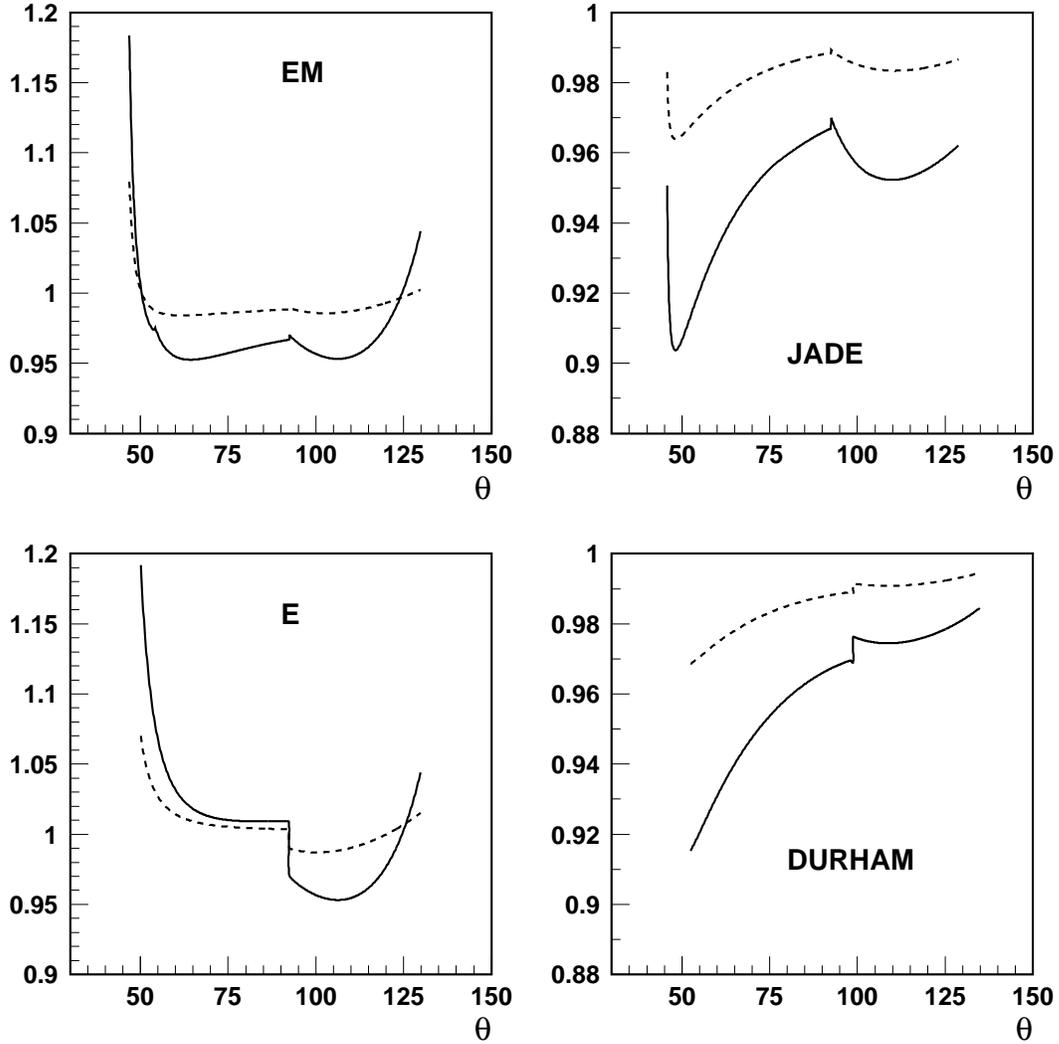
{The ratios of angular distributions $R^{bd}_\vartheta$
(see \protect\eq{ratangdis})
for $\yc=0.04$ for
the different algorithms. Solid line for  $m_b=5$~GeV and dashed line
for $m_b=3$~GeV.}
{ang}
\end{document}